\documentclass{mpe_report}
\usepackage{psfig}

\begin{document}
\title{Characteristic temperatures and spectral appearance of ULX disks}
\author{R.Soria\inst{1}, K.Wu\inst{1} \and Z. Kuncic\inst{2}}  
\institute{MSSL, University College London, Holmbury St Mary, Surrey RH5 6NT, UK
\and School of Physics, University of Sydney, NSW 2006, Australia}
\maketitle

\begin{abstract}
A standard disk around an accreting black hole may become 
effectively optically-thin and scattering dominated in the inner region, 
for high accretion rates (as already predicted by the Shakura-Sunyaev 
model). Radiative emission from that region is less efficient than blackbody emission, 
leading to an increase 
of the colour temperature in the inner region, by an order of magnitude 
above the effective temperature. We show that the integrated spectrum 
has a power-law-like shape in the $\sim 1$--$5$ keV band, with a soft 
excess at lower energies and a downward curvature or break at higher 
energies, in agreement with the observed spectra of many ULXs.
This scenario offers a physical explanation for the phenomenological  
``dense corona'', successfully used to fit those spectra. It also 
relates their fit parameters to physical properties of the accreting 
black hole, such as mass and accretion rate.

\end{abstract}

\section{Introduction}

Fitting colour temperatures to the X-ray spectra of 
accreting black holes (BHs) has traditionally been 
a useful, indirect method to estimate BH masses, 
typically within a factor of 2.
This method relies on the assumption that the accretion disk
extends down to the innermost stable circular orbit 
and is optically-thick, emitting a multicolour-blackbody spectrum 
(Shakura \& Sunyaev 1973; Frank, King \& Raine 2002).
If so, the fitted peak temperature and luminosity give a characteristic 
inner-disk area and hence BH mass. 
These conditions are usually verified in Galactic BHs 
in their high-soft state ($0.01 \la L_{\rm X}/L_{\rm Edd} \la 1$).

For ultraluminous X-ray sources (ULXs), 
the scenario that the disk is optically thick and reaches 
the innermost stable circular orbit could be a plausible 
initial assumption, in the absence of dynamic mass determinations. 
From a characteristic feature in the soft spectrum (``soft excess''), 
peak disk temperatures $\approx 0.15$--$0.20$ keV have been inferred and 
used to suggest masses $\sim 1000 M_{\odot}$ (Miller, Fabian \& Miller 2004). 
A strong objection against this argument is that most of the X-ray flux 
in ULXs comes out in a harder, power-law-like component, 
more consistent with inverse-Compton scattering
(Stobbart, Roberts \& Wilms~2006; Winter, Mushotzky \& Reynolds~2006). 
This suggests that the soft thermal component only traces emission 
from the outer, optically-thick disk, at radii $R \ga 10 R_{\rm ISCO}$; 
if so, its temperature and luminosity cannot be directly 
used to estimate the BH mass. An additional parameter 
is needed to model the location of the transition radius 
between the outer, standard disk and the inner, scattering-dominated 
region. In general, based on a variety of physical arguments, 
we expect that the transition radius 
$R_{\rm c} \propto \dot{m}^\beta R_{\rm ISCO}$ for $\dot{m} > 1$, 
where $\dot{m} \equiv \dot{M}/\dot{M}_{\rm Edd}$ and $\beta \sim 1$.


\section{Alternative physical interpretations}


A comparison of X-ray spectral and timing properties 
of ULXs and of Galactic BHs, in particular XTE J1550$-$564, 
has suggested a possible physical interpretation in terms 
of a standard disk plus Comptonizing corona (Goad et al.~2006; 
Done \& Kubota 2006). Unlike Galactic BHs, however, many ULXs have 
a second characteristic spectral feature at energies 
$\sim 5$--$10$ keV. There, the spectrum is not well approximated 
by a power-law: it shows a downward curvature or break 
(Stobbart et al.~2006; Roberts 2007). 
In the disk-corona framework, this can be interpreted 
as emission from a corona optically-thick to electron scattering 
($\tau^{\rm es} \sim 10$) and 
with a much lower temperature than is seen in Galactic BHs 
($kT \sim$ a few keV, compared with $kT \sim 100$ keV 
in Galactic BHs). In summary, the high-energy break 
would be related to the coronal temperature, and the soft component 
would come from the standard disk sticking out outside the corona, 
at large radii.

A possible difficulty of this model is that, in order to be optically thick, 
the corona must also be extremely dense. Assuming a characteristic 
radial size of $\sim 5000$ km, vertical size $\sim 1000$ km, 
$\tau^{\rm es} \sim 10$ corresponds to $n_e \sim 10^{17}$ cm$^{-3}$. 
This is many orders of magnitude denser than typical coronae 
in Galactic X-ray binaries. More importantly, one still has to explain 
how gravitational power is transferred from the underlying disk 
to the optically-thick corona. For proton densities 
$n_p \approx n_e \sim 10^{17}$ cm$^{-3}$, the thermal energy stored in 
such a corona is only $\sim 10^{35}$ erg, to be compared with 
a power $\sim 10^{40}$ erg s$^{-1}$ persistently transferred 
to the outgoing radiation via inverse Compton 
scattering. If gravitational power is released 
in the disk, a very efficient and stable energy transfer 
mechanism is required to keep the corona energized.


The same phenomenological spectral features 
have also been explained in a radically 
different way. The high-energy curvature and/or break have been 
interpreted as a very hot inner disk, with colour temperatures 
$kT_{\rm in} \approx 1.5$--$3$ keV: e.g., Stobbart et al.~2006).
A possible reason for such hot disks may be that the BH is rapidly 
spinning. Or, the disk may be non-standard, 
(e.g., slim disk models: Watarai, Mizuno \& Mineshige 2001).
In this scenario, ULXs are a natural extension of Galactic 
BHs, with higher accretion rates and hotter disks 
(Fig.~3 in Stobbart et al.~2006). 
The downside of this model 
is that it does not provide a natural explanation 
for the soft X-ray component. One possibility is that 
the soft excess comes from a downscattering outflow 
(King \& Pounds 2003) or from relativistically-blurred emission lines 
in the reflection spectrum of a photoionized disk 
(Crummy et al.~2006).

\section{Modified blackbody spectra}

In most ULXs, both scenarios (cool disk with a hotter Comptonised 
component, or a hot disk with a soft excess) are consistent 
with the observed spectra. Before we can use the peak 
disk temperature as a mass indicator, we need to determine whether 
the disk is associated with the soft or the hard  
components. Here, we suggest that both interpretations 
are in a sense correct, and follow simply from the standard 
disk model (Shakura \& Sunyaev 1973).

The thermal emission spectrum of an accretion disk 
is successfully approximated with a multicolor blackbody 
model, provided that two assumptions are verified: the disk 
is optically-thick in the vertical direction, 
and all the gravitational power dissipated during the infall 
is radiated.
Then (see, e.g., Frank et al.~2002 for details),
\begin{eqnarray}
I_{\nu}(R) &=& B_{\nu}[T(R)],\\
F(R) &=& \sigma T^4(R) \equiv \frac{3GM\dot{M}}{8\pi R^3}\,
\left[1-\left(\frac{R_{\rm ISCO}}{R}\right)^{1/2}\right],\\
F^{\rm obs}_{\nu} &=& \frac{2\pi \cos i}{d^2}
\int^{R_{\rm out}}_{R_{\rm ISCO}} I_{\nu}(R) \,R dR
\end{eqnarray}
where $D(R)$ is the power dissipated per unit disk area,
$d$ is the distance to the source, and $F^{\rm obs}_{\nu}$ is the observed flux.
Before substituting Eqs.~1 and 2 into Eq.~3, 
one has to make sure that the disk is optically thick at all $R$.
The effective optical depth (Rybicki \& Lightman 1979)
\begin{equation}
\tau^{\rm eff}_{\nu} \approx \sqrt{\tau^{\rm ff}_{\nu} (\tau^{\rm ff}_{\nu} + \tau^{\rm es})}
\approx (H/\cos i) \sqrt{\alpha^{\rm ff}_{\nu} (\alpha^{\rm ff}_{\nu} + \alpha^{\rm es})},
\end{equation}
where $H(R)$ is the disk thickness, 
takes into account both the (frequency-dependent) 
free-free absorption coefficient 
\begin{equation}
\alpha^{\rm ff}_{\nu} \approx 3.7 \times 10^8 \, T^{-1/2} Z^2 n_e n_i \nu^{-3} 
(1-e^{-h\nu/kT}) g^{\rm ff}\ {\rm cm}^{-1}
\end{equation}
(where the Gaunt factor $g^{\rm ff} \approx 1$), and the electron  
scattering coefficient 
\begin{equation}
\alpha^{\rm es} = n_e \sigma_T \approx 6.7 \times 10^{-25} n_e \ {\rm cm}^{-1}.
\end{equation}

At large radii, for $R \ga 7.2 \times 10^9$ $(\dot{m})^{2/3} (M/M_{\odot})$ cm  
(e.g., Frank et al.~2002) a standard disk is gas-pressure dominated, 
optically thick ($\tau^{\rm eff}_{\nu} \gg 1$), and the Rosseland mean 
absorption opacity $\tau^{\rm ff}_{\rm R} \gg \tau^{\rm es}$.
In this zone, $I_{\nu} = S_{\nu} = B_{\nu}$ and the spectrum emitted 
by each annulus is a blackbody.

Moving towards smaller radii, there is a region where the disk 
is still gas-pressure dominated and optically thick ($\tau^{\rm eff}_{\nu} \gg 1$) 
but the opacity is dominated by electron scattering 
($\tau^{\rm es}_{\nu} \gg \tau^{\rm ff}_{\nu} \gg 1$). 
Deep inside the disk, the radiation is still in thermal 
equilibrium with the matter ($S_{\nu} = B_{\nu}$), but 
the emerging spectrum from the disk surface is modified by scattering, 
such that the outgoing intensity 
\begin{equation}
I_{\nu} = \frac{2B_{\nu}}{1+\sqrt{(\alpha^{\rm ff}_{\nu} + \alpha^{\rm es})/\alpha^{\rm ff}_{\nu}}} 
\approx 2 \sqrt{\frac{\alpha^{\rm ff}_{\nu}}{\alpha^{\rm es}}} \,B_{\nu} \la B_{\nu}
\end{equation}
(Rybicki \& Lightman 1979). From Eq.~(7) one can directly calculate 
the integrated disk spectrum (see Section 4). It is often useful to 
introduce an approximation of Eq.~(7) that does not contain 
a frequency dependence, by using
the Rossland mean of the absorption coefficient
\begin{equation}
\alpha^{\rm ff}_{\rm R} \approx 1.7 \times 10^{-25} \, T^{-7/2} Z^2 n_e n_i g^{\rm ff}_{\rm R}\ {\rm cm}^{-1},
\end{equation}
where the frequency-averaged Gaunt factor $g^{\rm ff}_{\rm R} \approx 1$. Then, 
the total emitted flux from a unit area on the disk 
\begin{equation}
F(R) \approx \frac{4}{\sqrt{3}} \sqrt{\frac{\alpha^{\rm ff}_{\rm R}}{\alpha^{\rm es}}} \, \sigma T(R)^4, 
\end{equation} 
that is, for a given temperature, the disk is slightly less efficient 
than a blackbody at radiating the dissipated gravitational power. 
If we impose that the disk has to radiate the same power, the colour 
temperature at each radius must increase with respect to the effective 
temperature: 
\begin{equation}
T'(R) \approx 0.81 (\alpha^{\rm es}/\alpha^{\rm ff}_{\rm R})^{1/8} \, T_{\rm eff}(R), 
\end{equation} 
where $1 < (\alpha^{\rm es}/\alpha^{\rm ff}_{\rm R})^{1/8} \la 2$ 
as we will show later.

At smaller radii, 
$R \la 1.1 \times 10^8 \, \alpha^{2/21} \dot{m}^{16/21}$ $(M/M_{\odot})^{13/21}$ cm 
(Frank et al.~2002), radiation pressure dominates over gas pressure. 
In this region, the disk may become optically thin to true 
absorption ($\tau^{\rm ff}_{\nu} \ll 1$), 
but is always optically thick to scattering 
($\tau^{\rm es}_{\nu} \gg 1$) and most importantly, is still effectively 
optically thick ($\tau^{\rm eff}_{\nu} \gg 1$).  
As long as $\tau^{\rm eff}_{\nu} \gg 1$, Eqs.~7 and 9 are still valid, 
and the emerging spectrum is still a modified blackbody. 
Explicitly, for a pure Hydrogen gas, from Eqs.~5 and 6:
\begin{eqnarray}
\frac{\alpha^{\rm es}}{\alpha^{\rm ff}_{\nu}} & \approx & 1.8 \times 10^{-33} \, 
T^{1/2} n_e^{-1}  \nu^{3} (1-e^{-h\nu/kT})^{-1},\\
\tau^{\rm eff}_{\nu} & \approx & (H/\cos i) \sqrt{\alpha^{\rm es} \alpha^{\rm ff}_{\nu}} 
\ \approx  1.6 \times 10^{-8} \, (H/\cos i)
\nonumber\\
&& \times \, n_e^{3/2}  \nu^{-3/2} (1-e^{-h\nu/kT})^{1/2}.
\end{eqnarray}

Applying standard-disk solutions for the radiation-pressure dominated region 
(Shakura \& Sunyaev 1973; Frank et al.~2002), we can approximate
\begin{eqnarray}
T(R) & \approx & 2.3 \times 10^7 \alpha^{-1/4} m^{-1/4} r^{-3/4} \ {\rm K},\\
H(R) & \approx & 2.5 \times 10^6 \dot{m} \,m 
\left(1-r^{-1/2}\right) \ {\rm cm},\\
n_e(R) & \approx & 6.1 \times 10^{17} \alpha^{-1} \dot{m}^{-2} m^{-1} r^{3/2} \nonumber \\ 
&& \times \left(1-r^{-1/2}\right)^{-2} \ {\rm cm}^{-3},
\end{eqnarray}
where we have defined $m \equiv M/M_{\odot}$, $r \equiv R/R_{\rm ISCO}$.
Inserting Eqs.~13--15 into Eqs.~11 and 12, we obtain:
\begin{eqnarray}
\frac{\alpha^{\rm es}}{\alpha^{\rm ff}_{\nu}} & \approx & 
2.0 \times 10^5 \, \nu_{\rm 1 keV}^3 \, (1-e^{-h\nu/kT})^{-1}  \, \alpha^{7/8} \nonumber \\ 
         && \times m^{-1/8}  \, \dot{m}^2  \, r^{-15/8} \left(1-r^{-1/2}\right)^{2}\\
\tau^{\rm eff}_{\nu} & \approx & 2.3 \times 10^{-3} \, \nu_{\rm 1 keV}^{-3/2} \, (1-e^{-h\nu/kT})^{1/2} 
                \, \alpha^{-23/16} \nonumber \\
         && \times m^{-7/16} \, \dot{m}^{-2}  \, r^{39/16} \left(1-r^{-1/2}\right)^{-3} (1/\cos i).
\end{eqnarray}


For typical 
Galactic BHs in a high state, the innermost part of the disk 
falls into the radiation-pressure, electron-scattering-dominated, 
effectively-thick regime. For $M=10 M_{\odot}$, $R=2R_{\rm ISCO}$, 
$\dot{M} = \dot{M}_{\rm Edd}$, viscous parameter $\alpha = 0.1$, 
$h\nu = 1$ keV (so that $1-\exp{(-h\nu/kT)} \sim 1$ across the $0.3$--$10$ keV 
band), we see from Eq.~16 that
the colour correction $f_{\nu}(R) \equiv 0.81 (\alpha^{\rm es}/\alpha^{\rm ff}_{\nu})^{1/8}$ 
$\approx 2$, only weakly dependent on frequency (within the X-ray band), 
accretion rate, BH mass and radius. In fact, $1.5 \la f_\nu \la 2.5$ 
throughout the X-ray emitting regions, for the characteristic range 
of physical parameters typical of the high/soft state in Galactic BHs.
That is why $f_{\nu}(R)$ can be approximated with a constant 
``hardening factor'' $f \approx 1.7$--$2.5$ 
(Shimura \& Takahara 1995; Shafee et al.~2006; Shrader \& Titarchuk 2003).

The previous results depend critically on the condition $\tau^{\rm eff}_{\nu} \ga 1$.
From Eq.~17, using the same set of parameters suitable 
for Galactic BHs in a high state, we see that $\tau^{\rm eff}_{\nu} \approx 5$ 
in the innermost region, at the Eddington accretion rate; 
at larger radii, $\tau^{\rm eff}_{\nu} \gg 1$.  
So, the optically-thick condition is always verified 
for Galactic BHs, but the X-ray emitting region gets close to becoming 
effectively thin. A small increase in viscosity, 
BH mass and/or accretion rate above Eddington will lead 
to the formation of a radiation-pressure dominated, 
{\it effectively-thin region in the inner disk} 
($\tau^{\rm eff}_{\nu} \la 1$, with $\tau^{\rm ff}_{\nu} \ll 1$ 
and $\tau^{\rm es}_{\nu} \ga 1$). 
If that happens, Eq.~7 is no longer applicable.
Note that an increase of the accretion rate leads to a reduction 
of the effective optical depth in the radiation-dominated zone 
(perhaps counter-intuitively). This is essentially because 
$n_e \propto \dot{m}^{-2}$. Conversely, in the gas-pressure dominated 
disk, an increase in $\dot{m}$ leads to an increase in optical depth.

In the region where the disk becomes effectively thin, 
the specific intensity emitted by an annulus is 
\begin{equation}
I_{\nu} \approx \left(1-{\rm e}^{-\tau^{\rm eff}_{\nu}}\right) \, 
\sqrt{\frac{\alpha^{\rm ff}_{\nu}}{\alpha^{\rm es}}} \, B_{\nu} 
\approx \frac{\alpha^{\rm ff}_{\nu} H}{\cos i} B_{\nu} \ll B_{\nu}.
\end{equation}
If all the flux is radiated from each annulus, 
the temperature must increase even more dramatically 
than in the optically-thick regime.
The hardening factor is now 
\begin{eqnarray}
f_{\nu}(R) & \approx & \left(\alpha^{\rm ff}_{\nu} H/\cos i\right)^{-1/4} \nonumber \\
    &\approx& 21.0 \, \nu_{\rm 1 keV}^{3/4} \, (1-e^{-h\nu/kT})^{-1/4} \, (1/\cos i)^{1/4}
\nonumber \\
                && \times \, \alpha^{15/32} m^{7/32} \, \dot{m}^{3/4}  \, r^{-27/32} \left(1-r^{-1/2}\right)^{3/4}.
\end{eqnarray}


Let us consider a hypotetical ULX with a mass $m=100$ and an 
accretion rate $\dot{m} \ga 1$. Then, the inner disk 
is effectively thin for ``reasonable'' choices of $\alpha$,
and the transition radius between thin and thick regions is 
\begin{eqnarray}
r_c & \la & 12.1 \alpha^{23/39} m^{7/39} \dot{m}^{32/39} \left(1-r^{-1/2}\right)^{48/39} \, (\cos i)^{16/39}\nonumber \\
& \sim & {\rm a\ few\ \ } \dot{m}^{0.8}.
\end{eqnarray}
For $r \ga r_c$, the spectrum is a disk-blackbody, 
slightly modified by a hardening factor 
$\approx$ constant over the frequencies and radii relevant to the X-ray 
spectral observations, and most likely $\la 2$. 
For $r \la r_c$, the hardening factor increases sharply 
towards the innermost stable circular orbit, 
reaching $f_{\rm 1 keV} \approx 4.3 \, \dot{m}^{3/4} \sim 10$ 
in the brightest region of the disk. Although the effective optical 
depth is $< 1$ in this region, the optical depth 
to scattering alone 
\begin{equation}
\tau^{\rm es}  \approx 6.7 \times 10^{-25} \, \frac{n_e H}{\cos i} \approx 
\frac{r^{3/2}}{\alpha \dot{m} \left(1-r^{-1/2}\right) \cos i} \ga 10.
\end{equation}

\section{Observational predictions}

If a disk develops an effectively-thin inner zone, how would its  
spectrum differ from a multicolour blackbody?
To have a qualitative idea, let us consider for simplicity 
a frequency-averaged hardening factor $f(R)\ge 1$, such that 
$I_{\nu}(R) = f^{-4}(R) B_{\nu}[T(R)]$ and $T(R) = f(R)T_{\rm eff}(R)$. Then,
the observed flux can be obtained as usual by integrating 
the emission over all disk annuli (Eq.~3):
\begin{eqnarray}
F^{obs}_{\nu} &=& \frac{2\pi \cos i}{d^2}
\int^{R_{\rm out}}_{R_{\rm ISCO}} I_{\nu}(R) \,R dR \\
&=& \frac{4\pi h \cos i \, \nu^3}{c^2 d^2}  \int^{R_{\rm out}}_{R_{\rm ISCO}} 
\frac{R dR}{f^4(R) \left\{{\rm e}^{h\nu/[kf(R)T(R)]} -1\right\}}. \nonumber
\end{eqnarray}

\begin{figure}
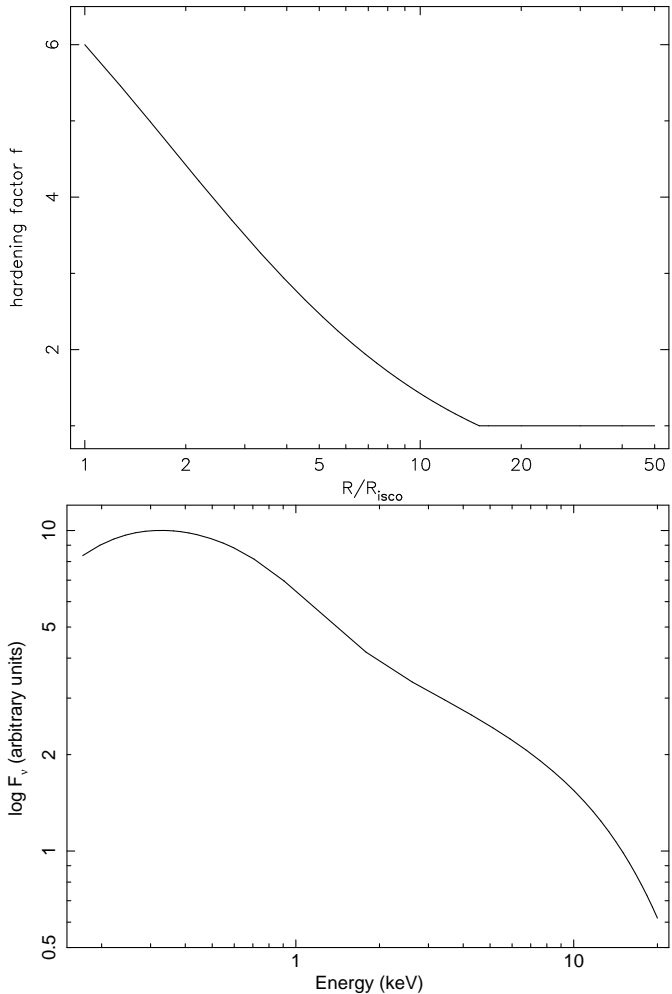

\centerline{\psfig{file=hardening.ps,width=8.8cm,angle=270} }
\centerline{\psfig{file=spectrum.ps,width=8.8cm,angle=270} }
\caption{A hardening factor that increases sharply at small radii 
(top panel), as we suggest is the case in ULX disks, 
produces an emitted X-ray spectrum (bottom panel) with 
three characteristic features: a power-law-like component, 
a soft excess at low energies, 
and a curvature or break at high energies.
\label{fig1}}
\end{figure}


Using for example the hardening factor plotted in Figure 1 (top panel), 
where we have assumed that $f = 1$ for $r>r_c$, 
and applying Eq.~22,
we obtain the X-ray spectrum plotted in the bottom panel of Figure 1.
A detailed study of the emitted spectra is left to a paper 
currently in preparation. Here we simply want to point out 
the main qualitative result: the spectrum has a power-law-like 
shape for energies $1.5 {\rm \ keV} \la E \la 6$ keV, a downward 
curvature and break above 6 keV, and a ``soft excess'' with 
a characteristic temperature $T_c \sim 0.2$ keV. The latter component 
corresponds to the disk-blackbody emission from the outer disk 
($R > R_c$), where $f \sim 1$. The high-energy 
break corresponds to the maximum colour temperature in 
the effectively-thin, scattering-dominated inner disk: 
$T \approx  fT_{\rm eff}$ 
$\approx 6 \, T_{\rm eff}$.
This result is already well known (Shimura \& Takahara 1995; 
Fig.~12 in Shakura \& Sunyaev 1973), 
but its implications for ULX studies are not well explored.
In between the two thermal features, the slope of the 
power-law-like section of the spectrum depends on how steeply 
the hardening factor increases towards the centre. 
A slightly steeper slope is obtained when the frequency dependence 
in the hardening function $f_{\nu}$ is also 
taken into account. Additional steepening of the spectrum results 
from non-radiative contributions (outflows, Poynting flux) 
to the cooling flux from the inner region. Increasing the range 
of values of the hardening factor (for example, by increasing the viscosity 
or the accretion rate) moves the soft excess and high-energy break 
below and above the observed X-ray band, respectively, 
creating an apparently ``pure power-law'' spectrum.

\section{Conclusions: standard disk or corona?}

We have shown that the accretion disk in BHs 
with $M \ga 10 M_{\odot}$, $\dot{M} \sim$ a few $\dot{M}_{\rm Edd}$
is expected to develop a scattering-dominated, effectively-thin inner region 
($R \la 10 \, R_{\rm ISCO}$), with a corresponding increse in the colour 
temperature, to compensate for the decreased emission efficiency.
The predicted spectrum is very similar to what is observed 
from many ULXs at $L_{\rm X} \sim 10^{40}$ erg s$^{-1}$.
Phenomenologically, those spectra have been successfully 
fitted with an optically-thick disk-blackbody emission 
from the outer disk, plus a dense scattering 
corona in the inner region, ``covering'' the inner disk. 
We have shown here that the characteristic scattering opacity, 
size and temperature observationally inferred for those 
``coronae'' are the same as predicted for a radiation-pressure 
dominated standard disk, for 
accretion rates and BH masses consistent with ULX parameters.
So, we argue that there is no need to postulate an ad-hoc corona 
covering or replacing the inner disk. It is the standard disk 
itself that becomes a hot, scattering-dominated environment 
at small radii, and produces the observed spectra.
The advantage of seeing it as the inner disk rather than a corona  
is that we do not have to worry about how energy is transferred 
from one to the other, or how their interface looks like.
Moreover, in the scattering disk scenario, the peak temperature, 
optical depth and emitted flux are not just empirical parameters: 
they are predictable functions of $M$, $\dot{M}$, $\alpha$, 
known from standard-disk models. This provides the tools 
to determine the physical properties of the accreting BHs 
in ULXs from their fitted spectral parameters, and relate their 
spectral evolution to changes in the accretion rate.
In this scenario, both the cool ($T \sim 0.2$ keV) and 
the hot component ($T \ga 5$ keV) come from the disk. 

\end{document}